\documentclass[11pt,a4paper,dvipsnames]{amsart}
\usepackage{amsthm}
\usepackage{graphicx}
\usepackage{caption}
\usepackage{subcaption}
\usepackage{subcaption}
\DeclareMathOperator{\Do}{do}
\usepackage{amsfonts}
\usepackage{graphicx}
\usepackage{framed}
\usepackage{amssymb}
\usepackage{mathrsfs} 
\usepackage{tikz}
\usepackage{bm}

\usepackage{xcolor}
\definecolor{cornellred}{rgb}{0.7, 0.11, 0.11}

\usepackage[colorlinks=true,urlcolor=RoyalBlue,citecolor=RoyalBlue,linkcolor=RoyalBlue,linktocpage,pdfpagelabels,
bookmarksnumbered,bookmarksopen]{hyperref}

\usepackage{amsthm}
\usepackage{graphicx}
\usepackage{caption}
\usepackage{comment}
\usepackage{subcaption}
\usepackage{geometry}
\makeatletter
\@namedef{subjclassname@2020}{\textup{2020} Mathematics Subject Classification}
\makeatother

\numberwithin{equation}{section}
\begin{document}
\title[Causal Equal Protection as Algorithmic Fairness]{Causal Equal Protection as Algorithmic Fairness}

\date{}

\author[M.\,Di Bello]{Marcello Di Bello}
\author[N.\,Cangiotti]{Nicol\`o Cangiotti}
\author[M.\,Loi]{Michele Loi}

\address[M.\,Di Bello]{Arizona State University \newline\indent975 S. Myrtle Ave, Tempe, AZ 85287, United States} \email{mdibello@asu.edu}

\address[N.\,Cangiotti]{ Politecnico di Milano \newline\indent via Bonardi 9, Campus Leonardo, 20133, Milan, Italy}\email{nicolo.cangiotti@polimi.it}

\address[M.\,Loi]{Algorithmwatch\newline\indent Linienstraße 13, 10178, Berlin, Germany}
\email{michele.loi@icloud.com}



\begin{abstract}
By combining the philosophical literature on statistical evidence and the interdisciplinary literature on algorithmic fairness, we revisit recent objections 
against classification parity in light of causal analyses of algorithmic fairness and the distinction between predictive and diagnostic evidence. We focus on trial proceedings as a black-box classification algorithm in which defendants are sorted into two groups by convicting or acquitting them. We defend a novel principle, causal equal protection, that combines classification parity with the causal approach.  In the do-calculus, causal equal protection requires that individuals should not be subject to uneven risks of classification error because of their protected or socially salient characteristics.  The explicit use of protected characteristics, however, may be required if it equalizes these risks.
\end{abstract}
\maketitle

\section{Introduction}

A defendant is on trial for sexual battery.
The incriminating evidence comprises the victim's testimony, but corroborating physical evidence is missing. The prosecution seeks to introduce instead evidence that the defendant committed similar acts in the past, grew up in a violent  household, and was the victim of abuse. Research shows that people who fit this profile are more likely to commit sexual battery than the general population \cite{randSexualAss2015}. While by itself the profiling evidence cannot suffice for a conviction, it corroborates the victim's testimony. Cases like this in which case-specific evidence is combined with profiling evidence are  not uncommon.\footnote{See, among others, McLean v. State, 934 So. 2d 1248 (Fla. 2006). For an example of combining crime-specific evidence with profiling evidence in a drug case, see U.S. v. Vue, 13 F.3d 1206 (8th Circuit, 1994).} But we feel uneasy about relying on profiling characteristics as evidence of guilt, even when they are subsidiary to the main incriminating evidence.\footnote{Evidence law in the United States contains restrictions against the admissibility of bad character evidence or prior bad acts. To be sure, the scholarly literature is divided \cite{redmayneBookCharcter2015}.} 

Two promising explanations for this uneasiness can be found in the philosophical literature. According to one line of argument,  reliance  on traits such as prior bad acts is unfair toward defendants who possess those traits.  Someone who did not commit sexual battery but happened to fit the traits in question would risk being wrongfully convicted for reasons that have little to do with what they did, but someone who did not have the same traits would not be subject to this risk. Such uneven imposition of the risk of error seems unfair \cite{di_bello_profile_2020}. 
Another line of argument relies on the distinction between diagnostic and predictive evidence. Traits such as prior bad acts are predictors of criminality, not evidence that a crime occurred. 
By contrast, trace evidence 
such as fingerprints left by the perpetrator or the recollection of an eyewitness are examples of diagnostic evidence. They are caused via a physical or cognitive mechanism by the facts of the crime \cite{Lee-Stronach2023, thomsonIndEv1986} or are couterfactually sensitive to these facts \cite{enoch_statistical_2012}. We feel less uneasy about relying on diagnostic as opposed to predictive evidence against a defendant.\footnote{It cannot be the statistical or quantitative nature of the evidence that drives our uneasiness. For statistics are often admitted at trial, such as statistics about the reliability of eyewitnesses in a case in which the incriminating evidence consisted primarily in a witness testimony against the defendant, or statistics about the rarity of a genetic profile in a case in which the main evidence against the defendant is a DNA match.}  

Both these explanations are promising, but they are insufficient by themselves. For one thing, the predictive/diagnostic distinction cannot be the whole story because allowing only diagnostic evidence at trial would be too restrictive. Predictive evidence often helps to understand what happened and why, especially when it accompanies diagnostic evidence. 
The fairness-based account also falls short because it  overgeneralizes.  Defendants with different traits will often be subject to different risks of wrongful conviction, but this need not always be unfair. For suppose the investigators recover footprints inside an  apartment that was burglarized, and the footprints correspond to shoe size US men 10. 
Relying on shoe size would not be unfair even though only an innocent defendant with shoe size US 10 could risk being wrongfully convicted in the case at hand. 
 
 To make progress here, our working hypothesis is that the argument about fair risk imposition and the predictive-diagnostic  distinction complement one another. This paper weaves them together in a formally and conceptually rigorous way.  We defend a principle that we call \textit{causal equal protection}. We run through  stylized examples and test different fairness principles against the moral intuitions the examples generate. Only causal equal protection makes the right predictions in all the test cases. We offer an account of the fairness of risk imposition and also clarify in what circumstances diagnostic or predictive evidence satisfy it. 

Our argument combines two strands of literature, the philosophical literature on profiling and statistical evidence, and the  literature on algorithmic fairness in computer science, law and philosophy. The two strands have not been in much dialogue despite many points of overlap.  
To keep the discussion manageable, we focus on trial proceedings and view them as a type of black-box classification algorithm in which defendants are sorted into two groups by convicting or acquitting them. This is an idealization of sort, but one that helps to sharpen the question we seek to address.  We leave open the question to what extent our argument generalizes to questions of algorithmic fairness in other domains beyond trial proceedings.

Here is a preview of what is to come. As an explication of the idea of fair risk imposition, our starting point is the principle of classification parity which requires that long-run false positive and false negative error rates be equalized across two relevant groups, usually groups defined by protected characteristics such as race or gender. This is one of the most discussed principles in the literature on algorithmic fairness, also known under different names, such as equalized odds or equality of opportunity \cite{barocas_fairness_book_2023,hardt_equality_2016}. It is an easily applicable and intuitive principle, but despite that, it has come under increasing scrutiny  \cite{Beigang23, loiEtAlFairEquality2023, Hedden21,  Long21, Mayson19, hellman_measuring_2019, corbett-davies_measure_2018}.  The problem is that classification parity is violated if group disparities in error rates arise even when they are due to mere accidental factors, and conversely, classification parity may fail to detect underlying biases in the decision-making process. So it is both too strong and too weak of a criterion of fairness (Section \ref{sec:class-parity}). 

As an alternative, the literature has recently proposed causal criteria of algorithmic fairness, for example, in terms of counterfactual judgments grounded in causal models \cite{kusner_counterfactual_2017} or causal paths  \cite{chiappa_path-specific_2018}.
On this approach, whenever a (morally objectionable) causal influence runs from a relevant group variable to the algorithmic classification, this would be ground for considering the algorithm unfair. This approach improves on classification parity: in absence of an objectionable causal influence, it would not count a classification algorithm unfair even when group differences in error rates exist  (Section \ref{sec:causal}). But, we argue, the causal approach still falls short 
because it neglects the distinction between predictive and diagnostic evidence, a distinction that has  gone largely unnoticed in the literature on algorithmic fairness 
(Section \ref{sec:diagnostic}). 

There is a kernel of truth in classification parity as a criterion of algorithmic fairness since group disparities in error rates often signal an underlying bias in the algorithmic classifications. What we need is an account that combines classification parity with the causal approach.  This is precisely what our principle of causal equal protection accomplishes. 
When  formulated in the do-calculus \cite{pearlCausality2009}, causal equal protection mirrors the simplicity of classification parity and requires that individuals  not be subject to uneven risks of classification error because of their protected or socially salient characteristics. 
(Section \ref{sec:causal-equal-protection}). The principle also allows us to draw a plausible distinction between objectionable and unobjectionable forms of legal evidence on grounds of fairness (Section \ref{sec:legal}).

\section{Classification Parity}
\label{sec:class-parity}

\subsection{An intuitive principle}

Classification parity requires that the rates of false positive and false negative classifications---the fraction of truly negative people who are falsely classified as positive and the fraction of truly positive people who are falsely classified as negative---be the same across two relevant groups.  
Even though the requirement of equality in the rates of error is usually limited to groups defined by protected attributes, it can also apply more broadly to socially salient groups defined by characteristics such as socio-economic status or education level. But the principle is not usually applied to any arbitrary group, a limitation we should keep in mind. This limitation is sensible, or else there would always be some group, no matter how artificially defined, for which classification parity is violated.

Since our focus is on trial proceedings, it is instructive to see some examples of how violations of classification parity signal an underlying unfairness in the decision-making process. 
Consider the sexual battery case we started out with. According to classification parity, relying on prior bad acts and other bad traits as incriminating evidence would be unfair if it resulted in a disparity in the rates of false positives---false convictions---across the relevant groups defined by protected or  socially salient characteristics. 
If abuse and household violence are more prevalent among disadvantaged families and racial minorities,
 using traits such as prior bad acts as evidence of guilt  would make it easier to convict defendants in those groups, even when innocent.\footnote{According to a report to the US Congress \cite{reportCongressChildAbuse2010}, `[c]hildren in low socioeconomic status households had significantly higher rates of maltreatment' (p.\ 14) and `[t]he incidence of children seriously harmed ... was 8.8 per 1,000 Black children compared to 4.6 per 1,000 White children' (p.\ 4-24).} Thus, the rates of wrongful convictions would be higher for disadvantaged families and racial minorities. 
 This seems unfair. 
 
 Here is an even more straightforward example. Suppose, as the data on exonerations suggest \cite{raceWrongConvict2022}, the rate of false convictions is higher among Black defendants than white defendants. Absent race-neutral explanations, this disparity signals an unfairness in how trial proceedings treat defendants of different races.

\subsection{Hedden, Long and Beigang}

Even though classification parity is an intuitive criterion of algorithmic fairness, counterexamples to it abound. Following \cite{Hedden21}, imagine a scenario in which each person is associated with their objective chance of committing a crime. Suppose an algorithm can faithfully track the objective chance of each person and assign an equivalent dangerousness score. 
Positive classifications (say convictions) are made if a person's dangerousness score is above a threshold, and otherwise the classification is negative (say an acquittal). Next, suppose people are sorted into two rooms, and it just so happens that the distribution of the scores (and thus of the objective chances) across the two rooms is different. 
For example,  people in the first room tend to have higher scores compared to people in the second room. Going through a bit of math, it  is easy to see that the algorithm's rates of false positives (false convictions) and false negatives (false acquittals) across the two rooms---so long as the same decision threshold is used---will often differ.\footnote{
Imagine, for instance, two rooms with $20$ people in each. In the first room, the dangerousness score is $.9$ for $10$ people and $.1$ for the other $10$. In the second room, $10$ people have a score of $.4$ and the other $10$ a score of $.6$. Assuming scores track objective chances, the base rate of criminality is the same across the two rooms: there are 10 innocent people in each room. If the decision threshold is fixed at $.5$, the false conviction rate in the first room is 1/10 (since only one innocent is convicted), whereas the the false conviction rates in the second room is 4/10 (since four innocents are convicted). So classification parity is violated.}
Although classification parity relative to room membership is violated, Hedden points out that this violation does not make the algorithm intuitively unfair. 


Another example by \cite{Long21} makes a similar point. Suppose student papers are graded on the basis of familiar criteria: grammar, originality, structure, etc.  These criteria are reliable yet fallible indicators of a paper's quality, say, whether the paper is a true A paper. Students are sorted into two sections, but the papers are graded using the same criteria no matter the section.  So whether the students belong to one section or another has no effect on how they are graded. But it just so happens that the papers in one section are overall of better objective quality: there are more true A papers in one section than in the other. As in Hedden's scenario, classification parity is violated across the two sections, but there seems to be nothing unfair toward the students in one section or another.


We share Hedden's and Long's intuitions, but the force of these counterexamples can be downplayed because the groups in question, room membership or college section, are socially inert \cite{LazarStone, viganoEtAlCoin2022} and are randomly assigned \cite{sogaardEtAlAgainstHedden2024}. What if these groups were socially salient and played causally identifiable roles in the processes that shape society? An example due to  \cite{Beigang23} can help here, suitably modified to fit the criminal justice context. Suppose a trial system uses `age' as a determining factor for convictions in violent crimes since data show that young people are more prone to violent crime than older people. And suppose that in one county, Christians tend to be younger than Muslisms, and in another county, they tend to be older because of different patterns of urbanization, migration and employment.\footnote{This set-up portrays a common situation: the same algorithmic classification system---a trial system---is deployed in different contexts, different counties with their own demographics.} Beigang  observes that classification parity is violated in both  scenarios across different religious groups 
 but in opposite directions. Other things being equal, false conviction rates among Muslisms will be higher in one county and lower in the other.
 If we take classification parity as a criterion of fairness, the trial system would be biased against Muslims \textit{and} biased against Christians.  That is counterintuitive since the trial system cannot be biased against both.
 



Beigang's case portrays a violation  of classification parity relative to a socially salient and protected category, religion. 
But this violation
does not give rise to the intuition that the algorithm is unfair for that reason.\footnote{The algorithm may well be unfair for other reasons, for example, because it relies on age and discriminates between people of different ages.}  A plausible diagnosis of what is going on  is that the variables `religion' and  `age' used by the trial classification algorithm are merely correlated.  The correlation may be explained by a common cause such as migration, but whatever the case, religion does not  affect how trial classifications are made, even though classification parity is (systematically) violated relative to religion. Since religion plays no causal role in the classification process, no individual is judged at trial differently \textit{because of} their religion.

This diagnosis applies equally well to Hedden's and Long's examples. 
In Hedden's, the lack of causal influence of  room membership becomes apparent in this counterfactual: if any individual had belonged to one group (room) instead of the other, the algorithm would have assigned the same score and made the same decision about the individual. Individuals can change the room they belong to without changing the dangerousness scores assigned to them.\footnote{If the group in question had not been room membership but race, changing someone's race would probably change other characteristics of that individual. After all, race is embedded in complex ways in the structure of society. The same applies to gender and other protected categories. 
} 
In Long's scenario, if a student were to complain to the professor, their complaint would have no force. The professor could respond that, had they been assigned to the other section, they would have received the same grade. Section membership does not affect grading.

\subsection{Neither necessary nor sufficient}

So far we looked at classification algorithms that violated classification parity, even though our intuitions did not take them  to be unfair.  These cases show that classification parity isn't necessary for algorithmic fairness. What explains our intuitions---we suggested---is the absence of a causal influence running from an individual's protected or socially salient characteristic, such as race, gender, class, etc.\ to the algorithmic classification. To sharpen this  diagnosis, we lay out another counterexample to classification parity inspired by \cite{corbett-davies_measure_2018}. The example shows that, when an algorithm exhibits a clear group bias, classification parity may fail to detect it. So, it cannot be a sufficient criterion for algorithmic fairness either.

Suppose a jury has a clear bias against Black people. The fact that a defendant is Black makes the jury more inclined to convict the defendant all else being equal. In contrast to the jury's bias, prosecutors are biased in another way. They bring to trial Black individuals against whom there is  weak evidence, and instead white people are prosecuted only when there is  strong evidence of guilt.  Now, even though the jury (or the jury algorithm) convicts Black defendants more easily than white people, this bias does not become manifest in higher rates of false positives (false convictions) among Black innocent defendants. Since prosecutors bring to trial Black people even when there is weak evidence against them, 
a greater number of Black defendants are clearly innocent and this lowers the rate at which innocent Black defendants are convicted. At the same time, the fact that the jury is biased against Black defendants increases the rates at which Black defendants, including innocent ones, are convicted. By the combined effect of the jury's and the prosecutor's biases, classification parity is not violated.\footnote{
By Bayes’s theorem, $\mathbb{P}(C \vert I)$, the proportion of  innocent ($I$) who are convicted ($C$), equals $\mathbb{P}(C \vert I) = \mathbb{P}(I \vert C)/\mathbb{P}(I) \cdot \mathbb{P}(C)$. By increasing the proportion of innocent people $\mathbb{P}(I)$ and the proportion of convictions $\mathbb{P}(C)$ for Black defendants only (both by a factor $k$), holding $\mathbb{P}(I \vert C)$ fixed, the proportion $\mathbb{P}(C \vert I)$ of innocent defendants who are convicted  remains the same.
}  
But this trial system is clearly unfair. 
Individual Black defendants could legitimately complain that, had they been white, they would have been treated differently.  


What has gone wrong with classification parity? In Hedden's, Long's and Beigang's examples, the group characteristic---room, section membership or religion---had no causal influence on the algorithmic classification. In the biased trial system, by contrast, the group characteristic---race---clearly had a causal influence.  So classification parity falls short as it is sometimes indifferent to causal influence. In addition, in Hedden's, Long's or Beigang's examples, no individual could complain of being judged differently because of the room, section or religion they were in. In the biased trial system, each individual Black defendant could complain of being judged differently because of their race. 
In sum, our diagnosis of the problem is this: classification parity falls short in that (i) it does not track whether or not individuals can legitimately complain of unfair treatment and (ii) it does not track the extent to which the unfair treatment of the individual can be causally traced to a protected characteristic they possess. Our aim is to put forward a principle that avoids these shortcomings. The  natural next step is to consider existing causal analyses of algorithmic fairness.

\section{The Causal Analysis}
\label{sec:causal}

\subsection{Correlation v.\ causal influence}
In order to model formally the idea of causal influence (or lack thereof), causal graphs are helpful.\footnote{We will be working with Bayesian networks in which the arrows between nodes (variables) are understood as relations of direct causal influence. To this end, two conditions must be met, the Markov condition and minimality. The Markov condition is standard and requires that each node is probabilistically independent of its non-descendants conditional on its parents \cite{pearlCausality2009}. The minimality condition ensures that there are no excess or redundant arrows between nodes \cite{kinneyCausalNet2023}.}
The age-based algorithm in Beigang’s scenario can be represented by the causal graph in Figure \ref{Fig:age-religion}.
There is a causal arrow from the variable `age' to the classification because age is used as evidence of guilt, but there is no causal arrow between `religion' and `age'. The protected category, religion, does not cause someone to be a certain age, even though a correlation---even a robust one---may exist between the two. So, if a Muslim defendant was convicted in part because of their young age, had the defendant been of a different religion, their age would not have changed and the classification would not have changed either. Religion does not affect the classification.  

By contrast, consider a trial system that relied on traits such as low  household income and childhood exposure to violence to corroborate case-specific evidence of guilt, where these traits are strong predictors of criminality. Suppose, as is plausible, discrimination in hiring makes it harder for people in minority groups to secure well-paying jobs \cite{pager_sociology_2008}. In addition, economic disadvantage, segregation and lack of investments in poor neighborhoods cause greater exposure to violence among minorities \cite{neighborRraceViolence2022}. 
 This setting 
 can be represented by the causal graph in Figure \ref{Fig:income}. An arrow runs from the variable `minority status' to `income' and `exposure to violence`. The arrow models the causal processes of discrimination and underfunding described earlier. 
Another arrow runs from the predictors to the classification itself because, by design, the algorithm makes classifications about criminality by considering the relevant traits that are predictive of criminality.  
So, putting it all together, a causal process runs from minority status to the trial judgment of conviction or acquittal. Here  minority 
status causally affects the classification.

\begin{figure}[t]
\begin{subfigure}[b]{0.8\textwidth}
    \centering
\begin{tikzpicture}[scale=1,>=latex]
\draw[very thick] (0.5,1) rectangle (2.5, 2);
\draw[very thick] (3.5,1) rectangle (5.5, 2);
\draw[very thick] (6.5,1) rectangle (8.5, 2);
\draw[very thick] (9.5,1) rectangle (11.5, 2);
\node[scale=0.8,text width=4.3cm] at (2.5,1.5) {Religion};
\node[scale=0.8,text width=4.3cm] at (5.5,1.5) {Age};
\node[scale=0.7,text width=4.3cm] at (8.2,1.5) {Evidence\\ Assessment};
\node[scale=0.7,text width=4.3cm] at (11.2,1.5) {Trial\\ judgment
};
\draw[thick,<->,dotted](2.5,1.5) --(3.5,1.5);
\draw[thick,->](5.5,1.5) --(6.5,1.5);
\draw[thick,->](8.5,1.5) --(9.5,1.5);
\end{tikzpicture}
\caption{Correlation, no causal influence.}
    \label{Fig:age-religion}
\end{subfigure}
\vspace{0.75cm}

\begin{subfigure}[b]{0.8\textwidth}
    \centering
\begin{tikzpicture}[scale=1,>=latex]
\draw[very thick] (0.5,1) rectangle (2.5, 2);
\draw[very thick] (3.5,1) rectangle (5.5, 2);
\draw[very thick] (6.5,1) rectangle (8.5, 2);
\draw[very thick] (9.5,1) rectangle (11.5, 2);
\node[scale=0.8,text width=4.3cm] at (2.5,1.5) {Minority \\Status};
\node[scale=0.6,text width=4.3cm] at (4.9,1.5) {Income/exposure\\ to violence};
\node[scale=0.7,text width=4.3cm] at (8.2,1.5) {Evidence\\ Assessment};
\node[scale=0.7,text width=4.3cm] at (11.2,1.5) {Trial\\ judgement
};
\draw[thick,->](2.5,1.5) --(3.5,1.5);
\draw[thick,->](5.5,1.5) --(6.5,1.5);
\draw[thick,->](8.5,1.5) --(9.5,1.5);
\end{tikzpicture}
\caption{Indirect casual influence.}
    \label{Fig:income}
\end{subfigure}
\vspace{0.75cm}

\begin{subfigure}[b]{0.8\textwidth}
    \centering
\begin{tikzpicture}[scale=1,>=latex]
\draw[very thick] (3.5,1) rectangle (5.5, 2);
\draw[very thick] (6.5,1) rectangle (8.5, 2);
\draw[very thick] (9.5,1) rectangle (11.5, 2);
\node[scale=0.8,text width=4.3cm] at (5.4,1.5) {Minority \\Status};
\node[scale=0.7,text width=4.3cm] at (8.2,1.5) {Evidence\\ Assessment};
\node[scale=0.7,text width=4.3cm] at (11.2,1.5) {Trial\\ judgement
};
\draw[thick,->](5.5,1.5) --(6.5,1.5);
\draw[thick,->](8.5,1.5) --(9.5,1.5);
\end{tikzpicture}
\caption{Direct causal influence.}
    \label{Fig:direct}
\end{subfigure}
\caption{}
\label{Figure:threegraphs}
\end{figure}
\noindent

\subsection{Direct v.\ indirect causal influence}

Beside contrasting cases of causal influence and those of mere correlation, it is also worth drawing another distinction, one between direct and indirect causal influences. 
In the trial system that 
relied on income and exposure to violence,  membership in a minority group is not itself used as evidence of guilt at trial. 
But even though membership in a minority group does not have a direct effect on convictions, its influence is indirect. 
Cases of indirect casual influence differ from those of direct influence. The latter are the easy ones and can be modeled formally by drawing a direct arrow from the relevant group variable to the algorithmic classification as in Figure \ref{Fig:direct}. 

If a protected or socially salient characteristic were explicitly used in trial decisions and affected convictions, that would be obviously unfair (though we will see in Section \ref{sec:causal-equal-protection} that this is not quite right). If the characteristic in question had a direct effect on convictions, but was used implicitly or subconsciously, that would also be morally objectionable. 
But what about cases of indirect influence? Could the trial system that relied on income and exposure to violence as corroborating evidence of guilt count as fair towards minorities? 
The answer might seem positive at first. The strong probative value of the predictors is a good reason for admitting them as corroborating evidence of guilt. In addition, people from households with the same income level and with the same level of exposure to violence, whether in the minority or the majority, will be judged the same. 
But this observation should not lead one to conclude that reliance on these traits does not give rise to any fairness concerns.

Reasoning counterfactually brings the issue into clear focus. 
Consider an innocent minority defendant who was wrongly convicted in part because they  grew up in a low income household and had early exposure to violence. Had they not been a member of a minority group, they would have been more likely to grow up in a higher income household  and thus less likely to be wrongfully convicted. This seems unfair towards them. Another way to see the point is to observe that by relying on household income and exposure to violence, the trial system  amplifies an unjust social process of discrimination that gives members of minority groups fewer opportunities. The antecedent bias in income distribution and access to safety taints the seemingly unbiased trial system.\footnote{On a related point, see the idea of compounding injustice in \cite{Hellman23}.} 

%

Some might insist that the bias isn't in the trial system; it exists in society. Discrimination in employment and underfunding of minoritized neighborhoods is the problem, not the trial system. This argument has some plausibility, especially because it is not obvious to whom a minoritized individual should complain. Should they complain to those who designed the trial system or society at large? 
At the same time, those who design decision systems, such as the judiciary, should not be oblivious to the larger context of society. Since decision systems have lasting impact on people's lives---and trials clearly do---their designers have a moral responsibility to take this  context into consideration \cite{lee_formalising_2021}.

\subsection{Causal paths and counterfactuals}
The discussion of direct v.\ indirect causal influences has highlighted two ways in which the causal analysis  grounds claims of algorithmic (un)fairness. 
One focuses on morally objectionable causal paths \cite{chiappa_path-specific_2018}.  For example, if income is allocated as a result of racially discriminatory hiring, the causal path from protected category (minority group) to the predictor (income) would itself be morally objectionable. This would morally taint the trial system that relied on income. So the unfairness of the decision system is explained by its reliance on a morally objectionable causal path in which a relevant group characteristic is implicated.

The other option is to say that any causal path in which a relevant group characteristic---say, a protected or socially salient one---affects the output classification would be, at least prima facie, morally objectionable. The existence of the causal path would warrant a pro tanto claim of unfairness. One way to defend this claim is to rely on counterfactual judgments. 
Counterfactuals track when an individual was---and would have been---judged  differently because of their race, gender or other protected characteristic. The differential treatment because of race, gender, etc.---which 
reflects underlying relations of causal influence---would warrant calling an algorithm unfair \cite{kusner_counterfactual_2017}.

A difficulty for the first approach---the path approach---arises when no morally objectionable path can be identified. Suppose that, because of cultural or lifestyle preferences that are constitutive of someone's racial identity, people of a certain race decide not to engage in the ``rat race" in our society. Their income would be  lower. Since the causal path from race to income is no longer morally objectionable, should any income-based trial decision be considered fair? Such a conclusion does not obviously follow. The path approach is thus too permissive. 
By contrast, the second approach based on counterfactuals captures the intuition that an income-based trial system would still be unfair towards those who have decided, for cultural or identity reasons, to conduct a modest life that puts them in the lower percentiles of the income distribution. 
But we will soon see that this version of the causal account faces difficulties of its own. To see why, we will introduce the distinction between diagnostic and predictive evidence, our  next topic.

\section{Diagnostic Fairness}
\label{sec:diagnostic}

\subsection{Diagnostic v.\ predictive evidence}
The causal analysis so far was confined to the relationships between socially salient categories (race, religion), predictors (income, age, exposure to violence) and algorithmic classification (conviction, acquittal). We should now consider more explicitly the role of the fact-related or outcome variable (whether or not the individual is factually guilty). 
Two cases are worth distinguishing here. 
First, predictors such as age or income are contributing forward-looking causes of the outcome. A predictor can be thought of as predictive evidence about the outcome, where the arrow of causality goes likes this: `predictive evidence $\to$  outcome'. 
But the direction of causality can also be reversed. Call evidence  that is caused by the outcome \textit{diagnostic evidence}, represented by the graph `outcome $\to$  diagnostic evidence' \cite{Lee-Stronach2023}.

The general structure of an unfair predictor-based algorithm, now augmented with a node for the outcome, is given in Figure \ref{fig:predictive}. 
Taking the causal graph as a good representation of reality, a change to the value of the socially salient category 
triggers---causally---a downstream change to the values of the outcome, as well as a change to the value of the classification. 
By contrast, consider Figure \ref{fig:diagnostic}. 
Here the classification is made using diagnostic evidence, not a predictor. Think of a criminal trial in which evidence against the defendant, say bullet casings or footprints found at the crime scene, fit the defendant's gun model or match the defendant's shoe size. Suppose the outcome---the criminal act---caused the bullet or the footprints to be present at the crime scene, not the other way around.\footnote{Footprints or bullet casings can of course also materialize at the crime scene before the crime occurred. We address this point toward the end of the paper.} In addition, suppose that the outcome were causally downstream relative to a protected category. For example, as seen before, because of discrimination people of a certain race tend to have lower incomes and thus are more likely to commit criminal acts.

\begin{small}
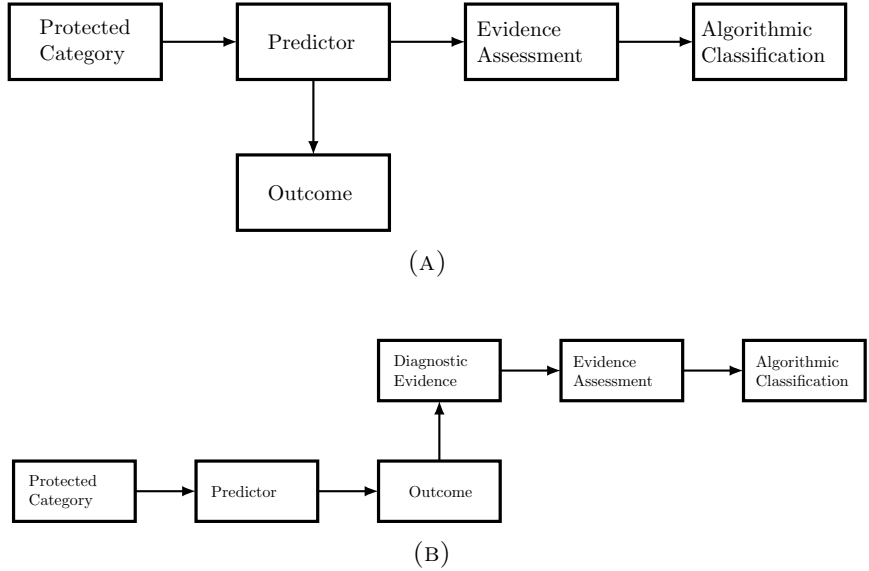
\begin{figure}[t]
    \centering
    \begin{subfigure}[t]{0.8\textwidth}
    \centering
\begin{tikzpicture}[scale=1,>=latex]
\draw[very thick] (0.5,5) rectangle (2.5, 6);
\draw[very thick] (3.5,5) rectangle (5.5, 6);
\draw[very thick] (6.5,5) rectangle (8.5, 6);
\draw[very thick] (9.5,5) rectangle (11.5, 6);
\draw[very thick] (3.5,3) rectangle (5.5, 4);
\node[scale=0.8,text width=4cm] at (2.5,5.5) {Protected\\ Category};
\node[scale=0.8,text width=4cm] at (5.5,5.5) {Predictor};
\node[scale=0.8,text width=4cm] at (8.2+3,5.5) {Algorithmic\\ Classification};
\node[scale=0.8,text width=4cm] at (8.25,5.5) {Evidence \\Assessment};
\node[scale=0.8,text width=4cm] at (5.5,3.5) {Outcome};
\draw[thick,->](2.5,5.5) --(3.5,5.5);
\draw[thick,->](5.5,5.5) --(6.5,5.5);
\draw[thick,->](4.5,5) --(4.5,4);
\draw[thick,->](8.5,5.5) --(9.5,5.5);
\end{tikzpicture}
 \caption{}
   \label{fig3a}
   \label{fig:predictive}
\end{subfigure}
\vspace{0.75cm}

\begin{subfigure}[b]{0.8\textwidth}
\begin{tikzpicture}[scale=0.8,>=latex]
\draw[very thick] (-0.5,3) rectangle (1.5, 4);
\draw[very thick] (2.5,3) rectangle (4.5, 4);
\draw[very thick] (5.5,5) rectangle (7.5, 6);
\draw[very thick] (8.5,5) rectangle (10.5, 6);
\draw[very thick] (11.5,5) rectangle (13.5, 6);
\draw[very thick] (5.5,3) rectangle (7.5, 4);
\node[scale=0.6,text width=4cm] at (4.25,3.5) {Predictor};
\node[scale=0.6,text width=4cm] at (1.25,3.5) {Protected\\ Category};
\node[scale=0.6,text width=4cm] at (7.25,5.5) {Diagnostic\\Evidence};
\node[scale=0.6,text width=4cm] at (10.2,5.5) {Evidence\\
Assessment};
\node[scale=0.6,text width=4cm] at (10.2+3,5.5) {Algorithmic\\ Classification};
\node[scale=0.6,text width=4cm] at (7.5,3.5) {Outcome};
\draw[thick,->](4.5,3.5) --(5.5,3.5);
\draw[thick,->](1.5,3.5)--(2.5,3.5);
\draw[thick,->](7.5,5.5) --(8.5,5.5);
\draw[thick,->](10.5,5.5) --(11.5,5.5);
\draw[thick,->](6.5,4) --(6.5,5);
\end{tikzpicture}
\caption{}
\label{fig:3b}
\label{fig:diagnostic}
\end{subfigure}
\caption{Algorithm based on predictive (\textsc{a}) v.\ diagnostic evidence (\textsc{b}).}
\end{figure}
\end{small}

The causal account from the previous section 
would make no difference between these two scenarios.
Since a causal path---even a morally objectionable one
---runs from the protected category to the classification, the counterfactual judgment would be the same: had the individual been of a different race, gender, religion, etc.\ they would have been judged differently, say more likely convicted than acquitted. This difference in trial judgment---due to a difference in protected or socially salient characteristic---would be the ground for the moral complaint by defendants in the relevant categories. 
%
%
%
The moral complaint could be voiced against algorithms based on (biased) predictive evidence (Figure \ref{fig:predictive})  as well as those based on (unbiased) diagnostic evidence (Figure \ref{fig:diagnostic}).  
But despite the causal dependency, we feel that the diagnostic evidence is used in the right way since the difference in judgment is due to a difference in the underlying facts. The causal account, then, overgeneralizes. 

\subsection{Biased diagnostic evidence}

To be clear, we are not saying that diagnostic evidence is immune from bias.  It is not. For imagine a society in which witnesses are racist: they are willing to testify against Black people but unwilling to testify against white people.  When they do decide to testify, they tell the truth, however. So eyewitness evidence is reliable when it can be obtained; it is simply hard to obtain it against white defendants. In this context, the availability at trial of reliable eyewitness evidence---a form of diagnostic evidence---counts as racially biased. Think of a causal arrow that goes from the protected category directly to the diagnostic evidence (Figure \ref{fig:diagnostic-biased}). Here trial decisions are racially biased because the witnesses whose testimonies are believed are themselves biased. 

Note the similarity between a classification algorithm that relies on racially biased witnesses and the predictive algorithm in Figure \ref{fig:predictive} that relies on a predictor such as income in a society in which hiring is racially discriminatory. Just like the biased witnesses, income as a predictor of criminality is biased since it is causally affected by the protected characteristic. 
By contrast, the algorithm in Figure \ref{fig:diagnostic} does not rely on biased diagnostic evidence, although the facts themselves are biased since they are causally influenced by the protected characteristic. 


\begin{small}
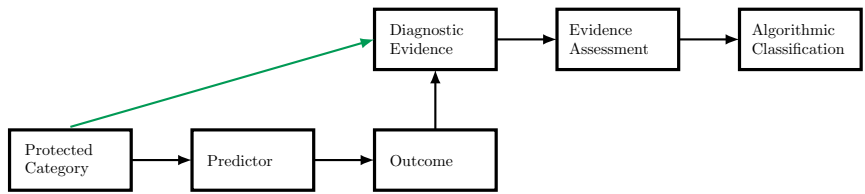
\begin{figure}[t]
    \centering
\begin{tikzpicture}[scale=0.8,>=latex]
\draw[very thick] (-0.5,3) rectangle (1.5, 4);
\draw[very thick] (2.5,3) rectangle (4.5, 4);
\draw[very thick] (5.5,5) rectangle (7.5, 6);
\draw[very thick] (8.5,5) rectangle (10.5, 6);
\draw[very thick] (11.5,5) rectangle (13.5, 6);
\draw[very thick] (5.5,3) rectangle (7.5, 4);
\node[scale=0.6,text width=4cm] at (4.25,3.5) {Predictor};
\node[scale=0.6,text width=4cm] at (1.25,3.5) {Protected\\ Category};
\node[scale=0.6,text width=4cm] at (7.25,5.5) {Diagnostic\\Evidence};
\node[scale=0.6,text width=4cm] at (10.2,5.5) {Evidence\\ Assessment};
\node[scale=0.6,text width=4cm] at (10.2+3,5.5) {Algorithmic\\ Classification};
\node[scale=0.6,text width=4cm] at (7.25,3.5) {Outcome};
\draw[thick,->](4.5,3.5) --(5.5,3.5);
\draw[thick,->](1.5,3.5)--(2.5,3.5);
\draw[thick,->](7.5,5.5) --(8.5,5.5);
\draw[thick,->](10.5,5.5) --(11.5,5.5);
\draw[thick,->](6.5,4) --(6.5,5);
\draw[thick,->,ForestGreen](0.5,4.05) --(5.5,5.5);
\end{tikzpicture}
\caption{Diagnostic evidence subject to two causal influences, from the outcome and  the protected category.}
\label{fig:diagnostic-biased}
\end{figure}
\end{small}

\subsection{Outcome-mediated paths}

To accommodate these nuances, the causal account can again invoke the distinction between morally objectionable and unobjectionable causal paths, following \cite{chiappa_path-specific_2018}. By stipulation, a path that contained the outcome variable would count as morally acceptable for the purpose of making assessments of algorithmic fairness. It is as though the fact that the outcome variable lies on the path overrides other claims of unfairness.\footnote{Such a path, as a whole, could still be morally problematic. If racist disinvestment in Black communities is a contributing cause of criminality, the disinvestment would  be morally objectionable for society and politics. 
But a trial system that made decisions on the basis of racially unbiased diagnostic evidence that tracked the facts 
should not be considered unfair toward Black defendants, even though the society in which it is embedded should.  We are drawing a distinction between fairness of classification systems, diagnostic or predictive, and fairness in how resources and investments are allocated, and policies are designed and implemented. 
}
So consider the following principle:

\begin{quote}
Any causal (directed) path from a protected or socially salient  group variable to the algorithmic classifications is  morally objectionable unless the fact-related variable of interest lies along the path. 
\end{quote} 

\noindent
Call this the \textit{diagnostic fairness principle}. 
That a Black defendant would have been judged  differently at trial had they been white---say, acquitted instead of convicted---need not always be morally problematic. Their complaint would have no force if it were based on the following consideration: had they been white, they would have been less likely to commit a crime and thus less likely to be arrested, tried and convicted.\footnote{Factually dissimilar people---factually guilty and factually innocent---should not be judged the same by the trial system; they should be judged differently. This is not to say that they should not be afforded equal procedural guarantees, such as the right to a defense or the right to be presumed innocent until proven guilty.} 
But their complaint would have force if it were based on this other consideration: had they been white, they would have been less likely to face incriminating evidence and thus less likely to be arrested, tried and convicted (just because of their race and not because of a difference in the underlying facts). The first, illegitimate complain is grounded in the causal graph in Figure \ref{fig:diagnostic}; the second, legitimate complaint in the causal graph in Figure  \ref{fig:diagnostic-biased}.

As a sanity check, note that the diagnostic fairness principle delivers the intended verdicts in  the other cases considered so far. There is no causal path from the protected category to the algorithmic classification in Hedden's, Long's or Beigang's scenarios. At best, there is a stable correlation. So the principle would not count the  algorithms in these scenarios as unfair.  Despite its emphasis on diagnostic evidence, the principle still leaves room for predictive evidence so long as it is unbiased, that is, not causally influenced by the group variable of interest. By contrast, the diagnostic fairness principle counts as unfair the income-based algorithm in Figure \ref{fig:predictive}  because of the causal path from the protected group variable to income, the predictor. All in all, the diagnostic fairness principle outperforms both classification parity and the existing causal analyses of algorithmic fairness. 
It is an attractive principle, but still not entirely satisfactory, as we shall see next. 

\section{Causal Equal Protection}
\label{sec:causal-equal-protection}

\subsection{Subtly biased diagnostic evidence}
 Consider a trial system that relied on diagnostic evidence, where the availability of this evidence was biased in a more subtle way than the racist witnesses who would only testify against Black people. Suppose that, because of racial segregation and redlining, Black people tend to live closer to high crime areas, and thus are seen around crime scenes more often than white people, even when they are innocent. Black innocent defendants will be more likely to be wrongfully convicted because of eyewitness testimony, even though the witnesses are not overtly racially biased. 
It is not objectionable to rely on such eyewitness testimonies against Black defendants under such circumstances even though a causal influence runs from the protected category `race' to the diagnostic evidence via racial segregation and redlining. 

This example depicts a situation in which, on one hand, a subtle racial bias exists which should not be ignored, but on the other, it would be odd to ban the use of eyewitness evidence in court for that reason.  A fix to this conundrum consists in adjusting the probative value of the eyewitness evidence to correct for the subtle eyewitness bias \cite{di_bello_profile_2020}. If someone's race influences, however indirectly and subtly, whether an eyewitness testifies against a defendant, 
the probative value of eyewitness evidence must be adjusted to compensate for this effect. In other words, the same eyewitness testimony will be assessed as having a different incriminating force depending on the race of the defendant.
How this fix works is best visualized in a causal graph (Figure \ref{fig:doublearrow}). First, a positive causal influence runs from the  category `race' to the diagnostic evidence: being of a certain race makes one more likely to face an incriminating eyewitness testimony. To neutralize this effect, adjusting the probative value of the diagnostic evidence results in a negative causal influence from the protected category `race' to the algorithmic classification (decision to convict or acquit).\footnote{Strictly speaking, the arrow goes first to the intermediate step called 
`evidence assessment'.}  Since the two causal influences have opposite signs, the overall causal influence of the protected or socially salient category onto the classification would be null.

\begin{small}
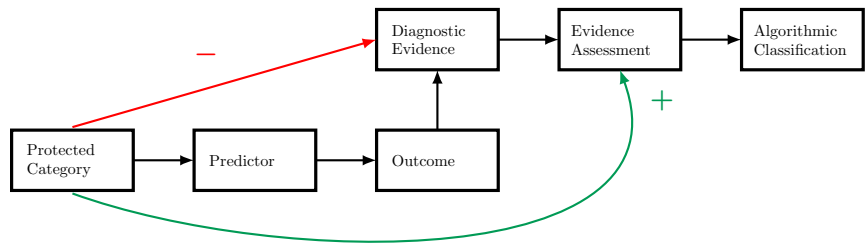
\begin{figure}[!ht]
    \centering
\begin{tikzpicture}[scale=0.8,>=latex]
\draw[very thick] (-0.5,3) rectangle (1.5, 4);
\draw[very thick] (2.5,3) rectangle (4.5, 4);
\draw[very thick] (5.5,5) rectangle (7.5, 6);
\draw[very thick] (8.5,5) rectangle (10.5, 6);
\draw[very thick] (11.5,5) rectangle (13.5, 6);
\draw[very thick] (5.5,3) rectangle (7.5, 4);
\node[scale=0.6,text width=4cm] at (4.25,3.5) {Predictor};
\node[scale=0.6,text width=4cm] at (1.25,3.5) {Protected\\ Category};
\node[scale=0.6,text width=4cm] at (7.25,5.5) {Diagnostic\\Evidence};
\node[scale=0.6,text width=4cm] at (10.2,5.5) {Evidence\\ Assessment};
\node[scale=0.6,text width=4cm] at (10.2+3,5.5) {Algorithmic\\ Classification};
\node[scale=0.6,text width=4cm] at (7.25,3.5) {Outcome};
\draw[thick,->](4.5,3.5) --(5.5,3.5);
\draw[thick,->](1.5,3.5)--(2.5,3.5);
\draw[thick,->](7.5,5.5) --(8.5,5.5);
\draw[thick,->](10.5,5.5) --(11.5,5.5);
\draw[thick,->](6.5,4) --(6.5,5);
\draw[thick,->,red](0.5,4.05) --(5.5,5.5);
\draw[thick,->,ForestGreen](0.5,2.95) to [out=-20,in=-70]  (9.5,5);
\node[scale=1,text width=4cm] at (5,5.25) {\color{red}{$\bm{-}$}};
\node[scale=1,text width=4cm] at (12.5,4.5) {\color{ForestGreen}{$\bm{+}$}};
\end{tikzpicture}
\caption{The positive causal influence from the protected category onto the classification is neutralized.}
\label{fig:doublearrow}
\end{figure}
\end{small}

The diagnostic fairness principle, however, cannot explain why this arrangement would be  fair or morally acceptable. Perhaps the principle could be amended with the proviso that, if a causal influence is meant to correct another, it should not count as objectionable. But this amendment would be \textit{ad hoc}.
After all, adjusting the probative value of the testimony depending on the relevant group identity of the defendant seems objectionable: an outright causal influence would be forced from the protected category `race' to the algorithmic classification. We need a more robust justification for why such a race-conscious differential assessment of the value of eyewitness evidence should be allowed.

\subsection{Unmediated causal influence in the aggregate}

We propose that talks of  causal paths be replaced by talks of unmediated overall causal influence.  Consider this principle:

\begin{quote}
An algorithmic classification counts as fair whenever the unmediated overall causal influence from the protected or socially salient category of interest to the algorithmic classification is even, that is, it does not change by switching the values of the relevant group variable,
say Black versus white defendants. The overall causal influence is assessed in the aggregate across all paths unmediated by the fact-related or outcome variable. The algorithmic classification  counts as unfair whenever the overall, unmediated causal influence changes in magnitude after switching the social category of interest.
\end{quote}

\noindent
Call this the principle of \textit{causal equal protection} (a more formal statement to follow shortly). 

The principle captures our intuitions in all the scenarios considered so far. There is no uneven causal influence of the relevant group variable in Hedden's, Long's and Beigang's scenarios simply because in them the relevant group variable has no causal influence onto the algorithmic classification to begin with. But when there is such a causal influence, two questions are relevant: first, whether the causal influence is mediated  by the fact-related variable, and second, whether the overall causal influence is even for different values of the relevant group variable.
There are different possible combinations here:

\begin{itemize}
\item[(a)] The only causal influence is
unmediated and uneven (Figure \ref{fig:predictive}).
\item[(b)] The only causal influence is mediated (Figure \ref{fig:diagnostic}). 
 \item[(c)] The causal influence is partly mediated and partly unmediated, and because it is partly unmediated, it is uneven (Figure
\ref{fig:diagnostic-biased}). 
\item[(d)]  The causal influence is again partly mediated and partly unmediated, but this time the unmediated influence is neutralized, so overall the influence is even (Figure
\ref{fig:doublearrow}).
\end{itemize}

\noindent
This list of cases is not exhaustive. For example, imagine a scenario with unmediated causal influences that in the aggregate cancel each other (Figure \ref{fig:2predict}).

\begin{small}
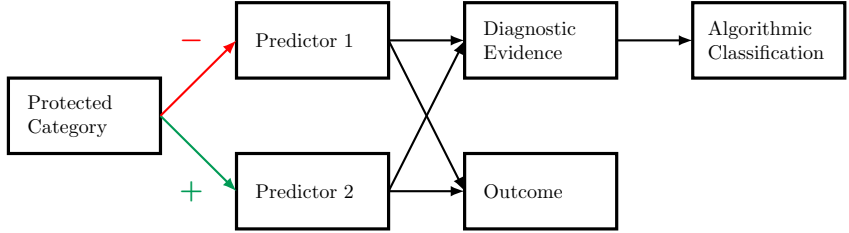
\begin{figure}[!ht]
    \centering
\begin{tikzpicture}[scale=1,>=latex]
\draw[very thick] (-0.5,4) rectangle (1.5, 5);
\draw[very thick] (2.5,3) rectangle (4.5, 4);
\draw[very thick] (2.5,5) rectangle (4.5, 6);
\draw[very thick] (5.5,5) rectangle (7.5, 6);
\draw[very thick] (8.5,5) rectangle (10.5, 6);
\draw[very thick] (5.5,3) rectangle (7.5, 4);
\node[scale=0.75,text width=4cm] at (4.25,3.5) {Predictor $2$};
\node[scale=0.75,text width=4cm] at (4.25,5.5) {Predictor $1$};
\node[scale=0.75,text width=4cm] at (1.25,4.5) {Protected\\ Category};
\node[scale=0.75,text width=4cm] at (7.25,5.5) {Diagnostic\\Evidence};
\node[scale=0.75,text width=4cm] at (10.2,5.5) {Algorithmic\\ Classification};
\node[scale=0.75,text width=4cm] at (7.25,3.5) {Outcome};
\draw[thick,->](4.5,3.5) --(5.5,3.5);
\draw[thick,->](4.5,5.5) --(5.5,5.5);
\draw[thick,->](4.5,5.5) --(5.5,3.5);
\draw[thick,->](4.5,3.5) --(5.5,5.5);
\draw[thick,->](7.5,5.5) --(8.5,5.5);
\draw[thick,->,ForestGreen](1.5,4.5)--(2.5,3.5);
\draw[thick,->,red](1.5,4.5)--(2.5,5.5);
\node[scale=1,text width=4cm] at (3.75,3.5) {\color{ForestGreen}{$\bm{+}$}};
\node[scale=1,text width=4cm] at (3.75,5.5) {\color{red}{$\bm{-}$}};
\end{tikzpicture}
\caption{Two unmediated causal influences balance each other out.}
\label{fig:2predict}
\end{figure}
\end{small}

\subsection{Even causal influence and fair risk imposition}

Besides vindicating intuitions in familiar cases, a more fundamental question remains.  Why is the overall unmediated causal influence of the relevant group category onto the algorithmic classification  morally relevant for questions of algorithmic fairness? The answer lies in the tight connection between the idea of even causal influence and that of fair risk imposition.  When the overall unmediated causal influence of a relevant group variable onto the classification is even, the risk of conviction is not unevenly distributed relative to that group variable. For example,  if the causal influence of the protected variable `race'  onto the classification is even, the risk of conviction is not unevenly distributed  among those who belong to different races. What matters for algorithmic fairness is not whether certain causal paths individually are morally objectionable, but whether all unmediated paths in the aggregate, whatever they are, ensure that innocent defendants in a certain category of interest are not subject to greater risks of false conviction than  others.

The connection between even causal influence and fair imposition of risk becomes  clearer if causal equal protection is formulated in the language of the $\Do$-calculus \cite{pearlCausality2009}:
\[P(C=1 \vert \Do[A=1] \text{ \& } \Do[Y=0]) 
= 
P(C=1 \vert \Do[A=0] \text{ \& } \Do[Y=0]),\]
where $C=1$ stands for a positive binary classification and the protected or socially salient binary group variable $A$ can take values $1$ or $0$. The variable $Y$ as is customary stands for the outcome, for example, the defendant is innocent ($Y=0$). The above formula should be understood following the idea of interventions originally developed by \cite{Pearl95}. Formally, the $\Do[...]$ operator updates the causal model by cutting the arrows incoming into the node and assigning a specific value to it, 
keeping the rest of the model unchanged.
\footnote{The mathematical computation of interventions strictly depends on the structure of causal graph \cite[Chapter 3]{PGJ16}.}

The causal influence of the relevant group variable $A$ is measured by the difference in the probability (risk) of a positive classification $C=1$ under the settings $\Do[A=1]$ versus $\Do[A=0]$. If the difference is zero,  the group variable $A$ has no uneven causal influence, or in other words, the risk is not unevenly allocated. 
The condition $\Do[Y=0]$---say the defendant's innocence---ensures that the (possibly uneven) causal influence of the group variable $A$ mediated by the outcome variable $Y$ are disregarded. This formulation also parallels the simplicity of classification parity, usually defined as:
\[P(C=1 \vert A=1 \text{ \& } Y=0) 
= 
P(C=1 \vert A=0 \text{ \& } Y=0)\]
While simplicity need not be the  mark of truth, it is satisfying to see that the different examples we considered are brought together under a principle that is formally rather simple.

\section{Applying the principle to legal evidence}
\label{sec:legal}

Having spelled out our principle and tested it against stylized examples, it is now time to see how it can answer the question we started out with. Is reliance on profile evidence at trial unfair, and if so, in what way it is unfair? Recall the sexual battery case from the beginning. That people who fit a particular profile---say, grew up in a violent household and were themselves victim of abuse in childhood---are more likely to commit sexual violence in their adulthood seems relevant information in a sexual battery case against a defendant who fit such profile.  Yet, we feel uneasy about using this evidence, and the question is whether this uneasiness can be justified on  grounds of unfairness.

\subsection{Socially salient traits} 
\begin{figure}[h]
    \centering
\begin{tikzpicture}[scale=1,>=latex]
\draw[very thick] (0.5,1) rectangle (2.5, 2);
\draw[very thick] (3.5,1) rectangle (5.5, 2);
\draw[very thick] (6.5,1) rectangle (8.5, 2);
\draw[very thick] (9.5,1) rectangle (11.5, 2);
\draw[very thick] (3.5,0) rectangle (5.5, -1);
\node[scale=0.8,text width=4.3cm] at (2.5,1.5) {Socially \\ salient trait};
\node[scale=0.7,text width=4.3cm] at (5.2,1.5) {Exposure to \\ violence};
\node[scale=0.7,text width=4.3cm] at (8.2,1.5) {Defendant\\ fits profile};
\node[scale=0.7,text width=4.3cm] at (11.2,1.5) {Trial\\ judgement
};
\draw[thick,->](2.5,1.5) --(3.5,1.5);
\draw[thick,->](5.5,1.5) --(6.5,1.5);
\draw[thick,->](8.5,1.5) --(9.5,1.5);
\draw[thick,->](4.5,1) --(4.5,0);
\node[scale=0.8,text width=4cm] at (5.4,-0.5) {Outcome};
\end{tikzpicture}
\caption{A socially salient trait has an effect on the trial judgment via the mediation of the trait `exposure to violence'  as incriminating evidence.}
    \label{Fig:violence}
\end{figure}
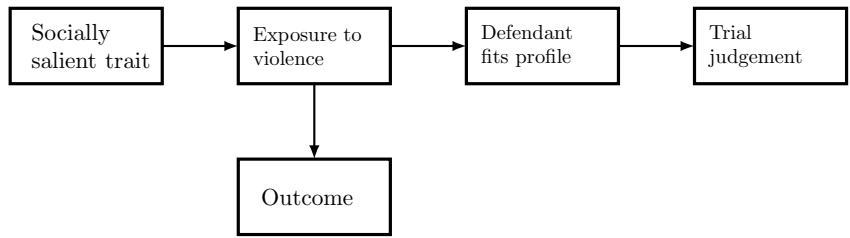

According to causal equal protection, fairness is violated whenever the defendant's socially salient traits, such as race, gender, or class have in the aggregate an uneven causal influence---direct or indirect---on the trial judgment.
If, for example, poverty affects one's early exposure to violence, then relying on a defendant's early exposure to violence as incriminating evidence would violate causal equal protection because being poor would positively affect the trial judgment and thus heighten one's risk of false conviction compared to, say, being wealthy. (See Figure     \ref{Fig:violence}, an instance of our graph in Figure  \ref{fig:predictive}.)

The fairness demands that causal equal protection imposes are limited to socially salient traits, and do not apply to any traits.  If there were no social processes that implicated class, race or gender in someone's exposure to violence,  this trait could be used as incriminating evidence without violating causal equal protection. So, one qualification to keep in mind is that the risk of wrongful conviction that should be equalized is understood as causal influence stemming from a socially salient trait of interest. A genuine question lingers here, namely what group traits should be selected. We have settled on protected characteristics and class, but of course the list could be extended or narrowed. This is ultimately a decision that must be informed by normative considerations and matters of policy.\footnote{\cite{di_bello_profile_2020} argue that, if such profiling evidence were used,  defendants who fit the profile but are otherwise innocent would be subject to a greater risk of wrongful conviction compared to those who do not fit the profile and are also innocent. The uneven allocation of the risks of wrongful conviction seems unfair. Their principle of equal protection requires that the risk of wrongful conviction not be higher for some innocent defendants compared to others facing comparably serious charges. But people with different traits---physical, temperamental, genetic, or what have you---will be subject for one reason or another to different risks of wrongful conviction should they face trial. So equal protection simpliciter seems too broad.}   



\subsection{Mediated causal influence}
Consider now the other challenging test case we started with. After examining the crime scene, an expert testifies that only people with shoe size US 10 could have left the footprints at the scene. Having that particular shoe size becomes incriminating evidence against those and only those who have that shoe size. 
%
%
Would using shoe size US 10 as incriminating evidence violate causal equal protection? 

Someone might answer affirmatively by reasoning as follows: an innocent defendant with shoe size US 10 would be subject to a higher risk of wrongful conviction compared to someone  with another shoe size since having a shoe size other than 10 could not be used as incriminating evidence. This observation by itself does not show a violation of causal equal protection insofar as shoe size is not causally affected by a socially salient trait. 
But it is easy to see that socially salient traits are likely to affect someone's shoe size, for example, men will tend to have larger shoe sizes. Quite plausibly, a socially salient trait would indirectly have an uneven causal influence on the trial judgment if shoe size were used as incriminating evidence. Causal equal protection would then seem violated. 

Here is an even simpler case. A witness reports having seen a Black man at the crime scene at the relevant time. Since this testimony can only incriminate Black men, race and gender would seem to have a direct, uneven causal influence on the trial judgment. Once again, causal equal protection would seem violated. The problem is that relying on someone's race or shoe size as identifying traits that match the perpetrator seems  unobjectionable, at least assuming that whoever is testifying about these identifying traits is not already biased against people of a certain race or gender in other ways. So we have a counterexample to causal equal protection---or do we really?  As we shall now see, a closer analysis shows that causal equal protection is, in fact, not violated in the test cases we are considering.

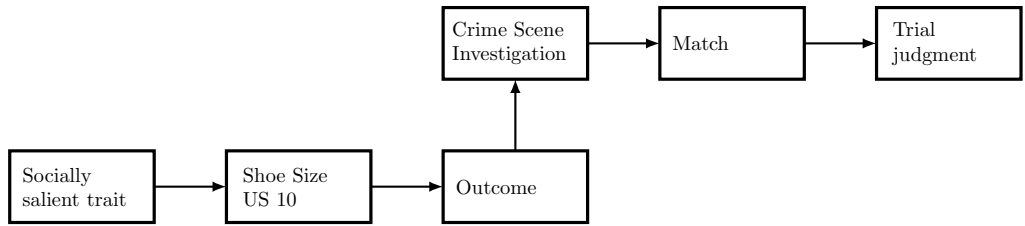
\begin{figure}[ht]
    \centering
\begin{tikzpicture}[scale=0.95,>=latex]
\draw[very thick] (-0.5,3) rectangle (1.5, 4);
\draw[very thick] (2.5,3) rectangle (4.5, 4);
\draw[very thick] (5.5,5) rectangle (7.5, 6);
\draw[very thick] (8.5,5) rectangle (10.5, 6);
\draw[very thick] (11.5,5) rectangle (13.5, 6);
\draw[very thick] (5.5,3) rectangle (7.5, 4);
\node[scale=0.7,text width=4cm] at (4.2,3.5) {Shoe Size \\ US $10$};
\node[scale=0.7,text width=4cm] at (1.15,3.5) {Socially\\ salient trait};
\node[scale=0.7,text width=4cm] at (7.1,5.5) {Crime Scene\\
Investigation};
\node[scale=0.7,text width=4cm] at (10.15,5.5) {Match};
\node[scale=0.7,text width=4cm] at (13.2,5.5) {Trial\\ judgment};
\node[scale=0.7,text width=4cm] at (7.15,3.5) {Outcome};
\draw[thick,->](4.5,3.5) --(5.5,3.5);
\draw[thick,->](1.5,3.5)--(2.5,3.5);
\draw[thick,->](7.5,5.5) --(8.5,5.5);
\draw[thick,->](10.5,5.5) --(11.5,5.5);
\draw[thick,->](6.5,4) --(6.5,5);
\end{tikzpicture}
\caption{Having a particular shoe size is incriminating only after gathering \textit{post hoc} information about the crime.}
    \label{Fig:match}
\end{figure}

The traits `shoe size US 10' or `Black man' become incriminating against a defendant who matches them only insofar as they are paired with information about what happened at the crime scene, for example, information that a crime was committed in a time and place and information that the perpetrator allegedly had those traits. The incriminating evidence, then, is not just the fact that someone is a Black man or has shoe size US 10, but rather, that there is a match between a trait associated with the crime and a trait of the defendant. 
To model this inferential structure, the match evidence can be represented in a causal graph as a node that lies downstream to the facts of the crime, or the outcome variable. In addition, the identifying traits such as shoe size or race can be placed upstream relative to the outcome variable, as in the graph in Figure \ref{Fig:match}, a modification of the earlier graph in Figure \ref{fig:diagnostic}. This is not say that having these traits makes one \textit{ex ante} more likely to commit a crime of the type in question---clearly, it does not. Rather, given case-specific \textit{post hoc} information about how, where and when the crime occurred, having such traits makes one more likely to commit the crime in question with the specific features that were discovered after the fact.\footnote{The graph in Figure \ref{Fig:match} incorporates case-specific information about the crime in question but the arrows should still be interpreted causally as we have done before.  That is, having shoe size US 10 makes one more likely to commit a crime in which the perpetrator leaves footprints matching size US 10. On the distinction between causal and epistemic interpretation of the arrows, see \cite{dahlman2022}.}




Crucially, the fact that having an identifying trait such as `shoe size 10' can be caused by protected traits such as race or gender no longer raises a fairness concern. Relying on a  match with `shoe size US 10' as incriminating evidence---and the same can be said of the trait `Black man'---does not violate causal equal protection because these traits do not exercise an uneven \textit{unmediated} causal influence on the trial judgment. Their causal influence is mediated by the outcome variable. Since the outcome variable mediates the causal influence of socially salient traits, causal equal protection is not violated. 
It is instructive to compare profiling evidence about exposure to violence (Figure     \ref{Fig:violence}) and match evidence (Figure     \ref{Fig:match}). In the former case, no \textit{post hoc} information about where, when or how the crime occurred is needed for profiling evidence to be incriminating against those who fit the profile. The only information needed is the type of crime in question. In the latter case, that a defendant matches certain identifying traits becomes incriminating evidence only conditionally on associating the same traits to the crime after a crime scene investigation has been conducted. 

\section{Conclusion}




We started out by contrasting two types of corroborating evidence: profiling characteristics such as prior bad acts or childhood exposure to violence and abuse, 
and evidence that the defendant matched an identifying characteristics of the perpetrator such as shoe size.  We argued that relying on them as evidence of guilt would be objectionable insofar as protected or socially salient features, such as race, gender or class  were to unevenly influence the trial judgment, in the aggregate and along paths not mediated by the outcome variable. Using some corroborating evidence or other would be unfair if defendants were subject to different risks of wrongful conviction because of their race, gender or social status.
This is the requirement of causal equal protection, which is essentially a causal interpretation of the well-known principle of classification parity in the literature on algorithmic fairness. 

Before we end, a clarification and a question are in order. First, the clarification: the assessment of risk imposition rests on an understanding of the relevant causal processes at work which are represented by arrows in the causal graphs. The causal processes that implicate the group variables of interest, such as race, gender or class, are typically systemic and operate at the level of society. In fact, socially salient categories, almost by definition,  are those that play a role in the relevant social causal processes. But the arrows in a causal graph can also represent how the decision-making process is structured, for example, that trial decisions are based on an assessment of the evidence, or how physical processes bring about evidence, for example, that the perpetrator's actions left physical traces at the scene. The assessment of the risk of wrongful conviction requires taking into account these different types of causal processes, social, physical and institutional.

Now 
the question: what if, as some have argued \cite{hu_whats_2020},  protected characteristics such as race or gender were not individual causes and could not be modeled as nodes in a causal graph? Admittedly, race and gender are complex, system-level social processes, not individual-level variables \cite{haslangerSocConstr2012}. We can only offer a brief sketch of an answer.  Causal graphs allow for relevant group characteristics to be modeled by the structure of the graph itself, not single nodes. Manipulations of group characteristics then would be manipulations of graph structures  \cite{malinskiIntervStruct2018}. Causal equal protection would then need to be reformulated, but the details belong to another paper.\footnote{Let $Y\leftarrow E\rightarrow C$ be a causal model, where $E$ is the evidence, $C$ the classification and $Y$ the outcome. The relevant group characteristic $A$ is not represented by the individual nodes, but by the causal coefficients between the nodes, for example, coefficient $\alpha_{EY}$ from $E$ to $Y$. Intervening on the group characteristic $A$ would mean to manipulate the value of the relevant coefficient $\alpha_{EY}$. Causal equal protection could be amended as follows (where $\alpha=x$ and $\alpha=x'$ are the different values of the group characteristic $A$ modeled by $\alpha$):
\[P_{\Do[\alpha=x]}(C=1 \vert \Do[Y=0]) 
= 
P_{\Do[\alpha=x']}(C=1 \vert \Do[Y=0])\]
}

\bibliographystyle{plain}
\bibliography{biblioJan2025}
\end{document}